\documentclass[aps,prl,twocolumn,groupedaddress,showpacs]{revtex4}
\usepackage{amsmath,amssymb}
\usepackage[dvips]{graphicx}
\newcommand{\beqar}{\begin{eqnarray}}
\newcommand{\eeqar}{\end{eqnarray}}
\newcommand{\beq}{\begin{equation}}
\newcommand{\eeq}{\end{equation}}
\newcommand{\ket}[1]{\left|{#1}\right\rangle}

\newcommand{\aver}[1]{\left\langle{#1}\right\rangle}
\newcommand{\added}[2]{\left|{#1},{#2}\right\rangle}
\newcommand{\inner}[2]{\left\langle{#1}|{#2}\right\rangle}

\begin{document}
\title{Ergodicity properties of quantum expectation 
values in entangled states}
\author{C. Sudheesh}
\email{sudheesh@prl.res.in}
\affiliation{Theoretical Physics Division, Physical Research Laboratory, 
Ahmedabad 380 009,  India}
\author{S. Lakshmibala}
\email{slbala@physics.iitm.ac.in}
\author{V. Balakrishnan}
\email{vbalki@physics.iitm.ac.in}
\affiliation{Department of Physics, Indian Institute of Technology-Madras, 
Chennai 600 036, India}
\date{\today}
\begin{abstract}
Using a model Hamiltonian for 
a single-mode electromagnetic field interacting with a nonlinear 
medium, we show that quantum expectation values 
of subsystem observables  can exhibit remarkably diverse ergodic 
properties even when the dynamics of the total system is regular.
The time series of the mean photon number is studied 
over a range of values of  
the ratio of the strength $\gamma$ of the 
nonlinearity to that of the inter-mode coupling $g$. 
We obtain the 
power spectrum, estimate the embedding dimension of the 
reconstructed phase space and the maximal 
Liapunov exponent 
$\lambda_{\rm max}$, and find the 
recurrence-time distribution of the 
coarse-grained dynamics.   
The dynamical behavior  
ranges from quasiperiodicity 
(for $\gamma/g \ll  1$)  to 
chaos as characterized by 
$\lambda_{\rm max} > 0$  (for  
$\gamma/g \gtrsim 1$),  and is interpreted. 

 \end{abstract}
\pacs{05.45.Mt, 42.50.-p, 42.50.Dv, 42.50.Md}
\keywords{Coherent states; photon-added coherent states; entanglement; 
time series; Liapunov  exponent; recurrence-time statistics.}
\maketitle

Classical nonlinear 
dynamical systems generically exhibit 
a long-time behavior that is ergodic---either on the 
energy surface (in 
Hamiltonian systems) or on some attractor (in dissipative
systems). A hierarchy of randomness can be identified, ranging from 
ergodicity to global exponential sensitivity to initial conditions. 
Quantum systems, on the other hand, are governed by linear
equations (the Schr\"odinger equation for the state vector
 or the  Liouville equation for the density operator).
 The randomness
in a quantum system, and
the precise manner in which its information content 
changes with time, are not easily determined. 

At least two different approaches have been 
employed in identifying signatures of 
the ergodicity properties 
of the quantum  counterpart of a generic classical 
system \cite{haake}. One of these relies on  the
observation that such signatures
are manifested in the energy-level 
statistics of the corresponding  quantum system.
If the system
is classically integrable, the quantum levels cluster together, 
and could even cross when a specific parameter in the Hamiltonian 
is varied \cite{berry1}.
A classically chaotic system, on the other hand, has its corresponding   
quantum levels so correlated as to resist such
 crossings \cite{berry2}.
Another approach
is based on the dynamics
of the overlap between two quantum states of the same physical                 
system which 
originate from the same initial
state, but with slightly different values of one of the control  
parameters \cite{peres}.
The time-dependent overlap is close to unity 
for all $t$  if the 
normalized initial 
state is located in a  regular region of the classical phase space. In
contrast, if the initial state is in a chaotic region of this 
space, the overlap falls off exponentially in time.

These lines of investigation concern quantum signatures of
classical dynamics. The inverse problem is of importance, and 
has also received attention: 
namely, the
identification of
signatures of non-classical effects 
such as wave packet 
revivals  in the temporal 
behavior of quantum expectation values (which, in turn,
could be regarded as effective dynamical variables 
in an appropriate `phase space').
(A (near-)revival of an initial  state  
$\ket{\psi(0)}$ 
 at $t = T_{\rm rev}$ implies that  
$|\inner{\psi(0)}{\psi(T_{\rm rev})}|^2 \simeq1$.)
The dynamics of a single mode of the radiation 
field governed by a nonlinear quantum 
Hamiltonian $H$ enables us to 
understand the connections between the 
behavior of quantum expectation 
values and various non-classical effects 
displayed  in wave packet dynamics \cite{sudh}.

In order to explore more thoroughly the 
range of dynamical behavior of
expectation values of observables, 
in particular in the presence of entanglement, 
we need a system in which revival phenomena can either occur or be 
suppressed, depending on the values of 
the parameters in $H$.
An uncomplicated but nontrivial  $H$ for our 
purposes is the one that  
describes the interaction of   
a single-mode field of frequency 
$\omega$  with the atoms of the nonlinear medium 
 through which 
it propagates. The medium is modeled \cite{agar1} 
by an anharmonic oscillator 
with frequency $\omega_0$ and 
nonlinearity parameter $\gamma$.
The Hamiltonian of the total system is given by
\begin{equation}
H = \hbar \omega \,a^\dagger a
+ \hbar \omega_0\,b^\dagger b + \hbar \gamma\,
b^{\dagger 2}\,b^2 +
\hbar g \,(a^\dagger b + b^\dagger a).
\label{2modehamiltonian}
\end{equation}
$(a, a^\dagger)$ are the field 
annihilation and creation operators, 
$(b, b^\dagger)$ are the
corresponding  atomic oscillator operators, and 
$g$ quantifies the coupling between the 
field and atom modes.
Importantly,   
although
the photon number operator $N = a^{\dagger}a$ does not commute with 
$H$ for any $g\neq 0$, 
the {\it total} 
number operator
$N^{\rm tot} = (a^{\dagger} a + b^{\dagger} b)$ does so 
for all values of the parameters in $H$.  While this 
implies that  $H$ can be cast in  block-diagonal form in a 
direct-product basis of field and atom Fock states, the model is not 
trivial. A simple way to see this is to re-write $H$ in terms of 
spin operators $J_i\,(i = 1,2,3)$ constructed from the two 
pairs of boson operators 
$(a, a^{\dagger})$ and $(b,b^{\dagger})$. It is then easily 
seen that the system of  
Heisenberg equations of motion for the spin operators 
does not close. 

The quantum mechanical expectation value 
$\aver{N(t)}$  (or the mean 
energy of the field mode)  
serves as a very convenient variable 
to probe the dynamics of this subsystem.
It varies with time because of the coupling between 
the two modes, and it deviates from periodicity because of the 
nonlinearity in $H$.  Its temporal behavior 
is remarkably diverse, ranging from 
quasiperiodic  to chaotic, depending strongly 
on the initial state and the parameter regime---in 
particular, on the degree of coherence of 
the initial state, and on the 
ratio $\gamma/g$ of the strengths of the 
nonlinearity and the field-atom coupling.  
We have studied the dynamics for states 
that are initially non-entangled
direct products of the field and
atomic oscillator states: specifically, states with
the field in a
 coherent state $\ket{\alpha}$\, (CS)  or 
an $m$-photon-added coherent state  $\ket{\alpha,m}$\,(PACS), 
while the atomic oscillator
 is in its ground state $\ket{0}$.   
Recall that the CS 
$\ket{\alpha}\,\,(\alpha \in \mathbb{C})$  satisfies
$ a\ket{\alpha}=\alpha\ket{\alpha}$, and 
is a minimum 
uncertainty state. The 
normalized PACS is defined \cite{agar2} as 
$\added{\alpha}{m}
=(a^\dagger)^m\ket{\alpha}/[
m!\,L_{m}(-\nu)]^{1/2}$  
where $m$ is a positive integer, $\nu = |\alpha|^2$,  
and $L_{m}$ is 
the Laguerre polynomial
 of order $m$.
A PACS possesses  
the useful properties of  quantifiable
and  tunable 
degrees of departure
from  perfect coherence and  Poissonian 
photon statistics. 
For brevity, we write
$\ket{\alpha} \otimes \ket{0} = \ket{\alpha\,;\,0}$ and 
$\ket{\alpha,m} \otimes \ket{0} = \ket{(\alpha,m)\,;\,0}$
for the initial states considered. $\aver{N(0)} = \nu$ and 
$[(m+1)\,L_{m+1}(-\nu)/L_{m}(-\nu)] - 1$, respectively,  
in the two cases.

We have carried out a detailed analysis of the
time series (using a time step 
$\delta t$ ranging from   
$10^{-2}$ for small 
$\gamma /g$ to $10^{-1}$ for large $\gamma /g$)
generated by the
values of the mean photon number $\aver{N(t)}$ 
computed over long intervals of time ($10^{6}$ time 
steps),  including  
phase space reconstruction, estimation of the
 minimum embedding
dimension $d_{\rm emb}$, calculation of the 
power spectrum \cite{abar,grass,fraser}, 
and recurrence-time statistics (using time series 
of $10^7$ steps when necessary). 
We use a robust algorithm
\cite{rosen} developed for
the estimation of the maximal Liapunov 
 exponent $\lambda_{\rm max}$ from
data sets represented by time series (see also \cite{kantz}).
The phase-space reconstruction 
procedure (including the extraction of 
$d_{\rm emb}$) 
has been carried out carefully, and 
it has been checked that  
any further increase in the dimensionality of the 
reconstructed phase space 
does not alter the inferences made regarding the 
exponential instability, if any, of the system. 
 
\begin{figure}[htpb]
\begin{center}
\includegraphics[width=3.4in]
{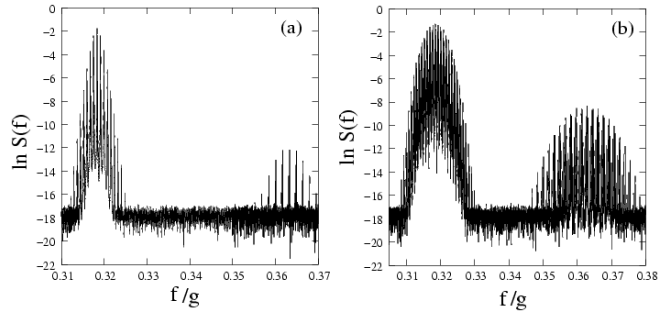}
\caption{Power spectrum of
the mean photon number {\it vs.} the frequency (in units 
of $g$) for the initial states
(a) $\ket{\alpha\,;\,0}$ and (b) $\ket{(\alpha,5)\,;\,0}$
with $\gamma/g=10^{-2}$ and $\nu=1$.}
\label{spectrumm0qbyg0.01nu1}
\end{center}
\end{figure}

For small values ($\ll 1$) of $\gamma/g$, 
near-revivals and fractional  revivals of the initial state occur, 
that are manifested 
in the entropy of entanglement of the system \cite{sudh4}.
 Correspondingly, we find that 
 the dynamics of 
$\aver{N(t)}$ 
ranges from periodicity to ergodicity, but
is not chaotic, essentially 
independent of the nature of 
$\ket{\psi(0)}$. 
As a case representative 
of weak nonlinearity, we have chosen
the parameter values $\gamma = 1,\,g = 100$.
Figure \ref{spectrumm0qbyg0.01nu1} (a), a
log-linear plot of the power spectrum $S(f)$ 
(the Fourier transform of the 
autocorrelation computed from the time series 
of $\aver{N(t)}$) as a function of the 
frequency $f$ when 
$\ket{\psi(0)} = \ket{\alpha\,;\,0}$,
is indicative of quasiperiodic behavior. 
With increasing lack of coherence
of the initial field state, the number
of frequencies seen in $S(f)$ increases. This
is already evident from
Fig. \ref{spectrumm0qbyg0.01nu1} (b), which
 corresponds to 
 $\ket{\psi(0)} = 
 \ket{(\alpha,5)\,;\,0}$.

\begin{figure}[htpb]
\begin{center}
\includegraphics[width=3.4in]
{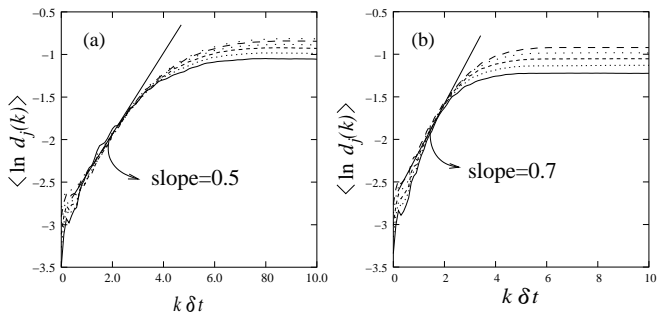}
\caption{$\aver{\ln\,d_j(k)}$ {\it vs.} $t$ for the initial states 
(a) $\ket{(\alpha,5)\,;\,0}$ 
with $\gamma/g=5$ and $\nu=1$ and
(b) $\ket{\alpha\,;\,0}$ with $\gamma/g=5$ and $\nu=10$.
The solid line corresponds to an embedding dimension 
$d_{\rm emb} =5$, and the  
dotted lines to values of 
$d_{\rm emb}$ 
from $6$ to $10$.}
\label{lyapm5and0nu1and10}
\end{center}
\end{figure}

In contrast to the case of weak nonlinearity,
the nature of the subsystem dynamics changes
drastically when $\gamma/g \gtrsim 1$.
As representative values 
for this nonlinearity-dominated regime, we  have set 
$\gamma = 5\,,\,\,g = 1$. 
We first examine the 
case corresponding to $\nu = 1$. For an initial field 
CS, both the time series 
and $S(f)$ 
confirm that the subsystem dynamics is not chaotic. 
In contrast to this,  
an initial PACS leads to a 
chaotic form for $S(f)$,  for sufficiently large values of $m$.
This is supported by an estimation of 
$\lambda_{\rm max}$ from the time series. 
The initial set of separations 
between the $j^{\rm th}$ pair of nearest 
neighbors in the reconstructed phase space 
evolves to the set  
$\{d_j(k)\}$ after $k$ time steps.  
Then $ \lambda_{\rm max}$ is the 
slope of the plot of  $\aver{\ln\,d_{j}(k)}$  
(the average is over all values of $j$) against $t$ 
in the linear region lying in between the  
initial transient and final saturation regions. 
Figure \ref{lyapm5and0nu1and10} (a) 
depicts
$\aver{\ln{d_j(k)}}$ {\it vs.}  $t$ for  
$\ket{\psi(0)} = \ket{(\alpha,5)\,;\,0}$, and 
the estimate of 
$\lambda_{\rm max}$, whose positivity  
indicates a chaotic variation of 
the mean energy of the 
field mode. 
For a given value of $\gamma/g$, an increase in  $\nu$
leads to chaotic behavior  even for an 
initial coherent field state.
This is demonstrated in 
Fig. \ref{lyapm5and0nu1and10} (b), which 
shows $\aver{\ln{d_j(k)}}$ {\it vs.}  $t$ 
for  $\ket{\psi(0)} = \ket{\alpha\,;\,0}$ with $\nu = 
10$. We further find that $\lambda_{\rm max}$ 
increases systematically with $m$ for an initial PACS.

Table 1 summarizes these conclusions.
\begin{figure}[htpb] 
\begin{center} \includegraphics[width=3.4in] 
{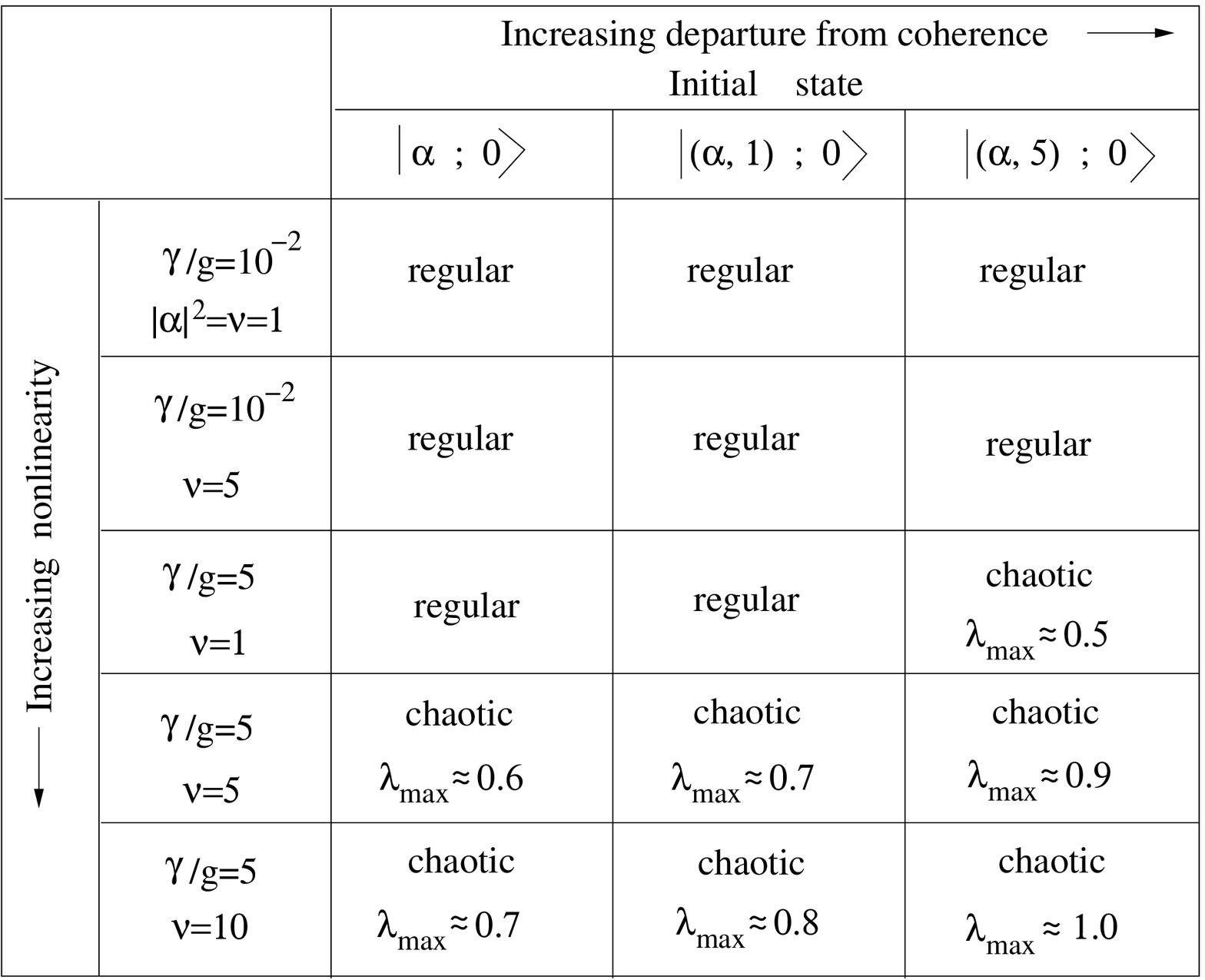}
\end{center}
Table 1: Qualitative dynamical behavior of the 
mean photon number of a 
single-mode electromagnetic 
field interacting with a nonlinear medium. ``Regular'' 
$\Rightarrow \lambda_{\rm max} = 0$. 
\end{figure}
In order to rule out  
round-off or truncation errors 
as the source of the 
computed chaotic behavior, 
we have 
verified in each case that the conlcusions are not 
altered if  $\aver{b^{\dagger}b}$ is chosen as 
the signal for which the time-series data is computed, 
and that $\aver{N} + 
\aver{b^{\dagger}b}$ does remain constant, as required.  The entropy of 
entanglement (not presented here) 
provides independent corroboration 
of the dynamical behavior as deduced from $S(f)$ and 
$\lambda_{\rm max}\,$. 

Our conclusions are reinforced by a detailed 
analysis of another important 
characterizer of dynamical behavior: recurrence statistics 
of the coarse-grained dynamics of 
$\aver{N(t)}$ 
as represented by its time series. 
For the range of
parameter values we use,  
the scatter in $\aver{N(t)}$ is typically $\gtrsim 1$.  We use a cell size 
$\sim 10^{-2}$ and very long time series. 
This enables us to numerically construct the invariant density 
$\rho$ (and hence the 
stationary measure 
$\mu$ for any cell $C$), as well as 
the distribution $F(\tau)$ of the 
time $\tau$ of first recurrence or return to $C$.
The mean recurrence time 
$\aver{\tau}$ 
can then be calculated, and 
compared with the result
$\aver{\tau} =  \mu^{-1}$ that follows from   
the Poincar\'e recurrence theorem.\cite{kac}. As the latter is 
derived from  the requirement of ergodicity alone, an  
agreement between the two values confirms  
that the dynamics is indeed ergodic in all the cases studied. 
We present here just two representative cases, 
both of which are also included in Table 1, for ready
reference. The first 
corresponds to weak nonlinearity 
($\gamma/g = 10^{-2}$) and  
$\ket{\psi(0)}= \ket{(\alpha,1)\,;\,0}, \, \nu = 1$. 
According to 
Table 1, this case is non-chaotic. 
\begin{figure}[htpb]
\begin{center}
\includegraphics[width=3.4in]
{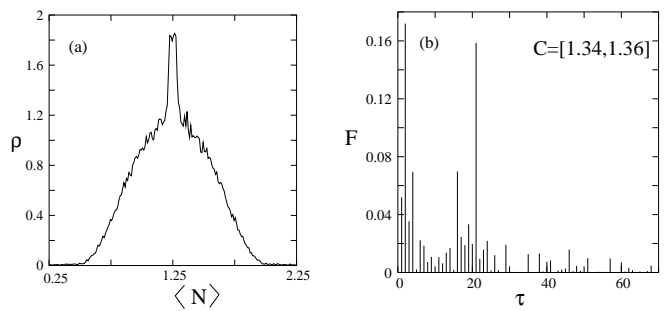}
\caption{(a) Invariant density and (b) first-recurrence-time distribution 
for the cell $C$ from the time series of $\aver{N(t)}$, 
for weak nonlinearity. $F(\tau)$ is characteristic of 
quasiperiodic dynamics.} 
\label{recm1gammabyg0.01nu1}
\end{center}
\end{figure}
Figure \ref{recm1gammabyg0.01nu1} (a) shows the invariant 
density, while (b) shows the actual recurrence time distribution. 
The discrete nature of the latter is a clear indication that the 
dynamics is actually quasiperiodic \cite{theu}. In marked contrast, 
consider a case of  strong nonlinearity,   
$\gamma/g = 5$ and 
$\ket{\psi(0)}= \ket{(\alpha,1)\,;\,0}, \, \nu = 10$.  
According to Table 1, 
this case is chaotic, with $\lambda_{\rm max} = 0.80\,$.  
\begin{figure}[htpb]
\begin{center}
\includegraphics[width=3.4in]
{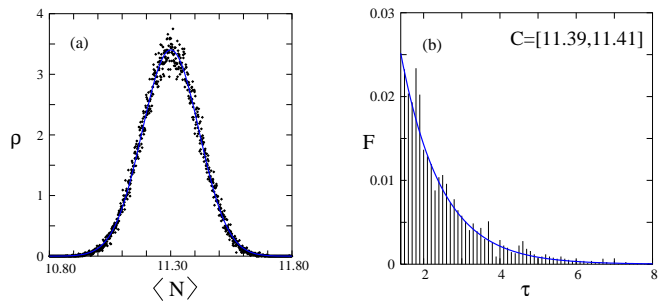}
\caption{(a) Invariant density and (b) first-recurrence-time distribution 
for the cell $C$ from the time series of  $\aver{N(t)}$, 
for strong nonlinearity. $F(\tau)$ is characteristic of 
chaotic dynamics.} 
\label{recm1gammabyg5nu10}
\end{center}
\end{figure}
Figure \ref{recm1gammabyg5nu10} (a) shows that the invariant density is in 
fact well-approximated by a Gaussian in this case. More importantly, 
$F(\tau)$ is very well fitted by the exponential 
distribution $\mu\,e^{-\mu \tau}$. This is precisely the 
distribution expected in a  hyperbolic dynamical 
system, for a sufficiently small cell size \cite{hira}.  
Moreover, 
in such a system successive recurrences to a cell 
must be uncorrelated and 
Poisson-distributed.  
We have further confirmed this feature by examining
the distribution of two successive recurrences, using  
even longer time series ($10^7$ steps). 
The distribution is again well fitted by the next term in the 
Poisson distribution, $\mu^2\tau\,e^{-\mu \tau}$. 

Now consider the completely 
{\it classical} counterpart 
of the Hamiltonian in Eq. (\ref{2modehamiltonian}).
Let the linear harmonic 
oscillator associated with $(a, a^{\dagger})$ have a mass $m$,  
position $x$ and momentum $p_x\,$, and let that  
associated with $(b, b^{\dagger})$ have a mass $M$, position $y$
and momentum $p_y\,$. Putting in all the constant factors (including  
$\hbar$) in the definitions of the raising and lowering operators
in Eq. (\ref{2modehamiltonian}), 
we get $H(x, \, p_x, \,y, \, p_y)$. 
When   $\hbar \rightarrow 0$,    
the only consistent way to obtain a non-trivial, finite 
expression for the classical Hamiltonian 
$H_{\rm cl}$ is to 
let $\gamma \rightarrow 0$ simultaneously, such that 
the ratio $\gamma/\hbar \rightarrow \lambda = $ a finite 
number. Then, with
$H_1 = p_{x}^2/(2m) + m \omega^2 x^2/2$ and 
$H_2 = p_{y}^2/(2M) + M \omega_0^2 y^2/2$,  
we find
\begin{eqnarray}
\hspace{-.5cm}
H_{\rm cl} & = &
H_1 + H_2 +
(\lambda/\omega_0^2)\,H_2^2 +\nonumber \\ 
& + & (g/\sqrt{\omega \omega_0})\left(
\sqrt{mM}\,\omega\omega_0\, xy + p_x \,p_y/\sqrt{mM}\right).
\label{classicalcoupledhamil1}
\end{eqnarray}
The counterpart of $N^{\rm tot}$ is  
$N_{\rm cl}^{\rm tot} = H_1/\omega + H_2/\omega_0\,$, 
which Poisson-commutes with  
$H_{\rm cl}\,$.
Hence the $2$-freedom classical system is
Liouville-Arnold integrable.  
Further,  although 
$H_{\rm cl}$ has 
cross terms that could 
change sign,  
$N_{\rm cl}^{\rm tot} = {\rm constant}$ 
is a hyperellipsoid 
in the $4$-dimensional phase space, so that
the motion is bounded 
for any set of initial conditions. All four Liapunov exponents
vanish, and the classical motion is always regular, and restricted  
to a $2$-torus, for each set of initial conditions. This behavior is
indeed very different from the much more diverse 
one found for the 
quantum expectation value  $\aver{N(t)}$.  

What, then,  is the 
interpretation of a positive value for 
$\lambda_{\rm max}\,$, 
and the hyperbolicity implied by the 
recurrence-time statistics, 
 as deduced from the time series for 
the subsystem variable $\aver{N}$ 
(equivalently, $\aver{b^{\dagger} b}$) 
in the quantum mechanical case? 
To start with, we note that 
a comparison of the quantum and classical 
cases is not always straightforward\cite{litt}, and  
the case at hand is one such instance.    
Letting $\gamma \rightarrow 0$ would remove the nonlinear 
term $b^{\dagger 2} b^2$ in the quantum $H$. 
In that case the dynamics of $\aver{N(t)}$ reduces to 
a trivial periodic exchange of energy 
between the field and atom modes.  
The origin of this 
dichotomy can be traced back to the inadequacy 
of the naive Ehrenfest 
theorem, which does not generally take into account 
the non-commutativity 
between 
$x$ and $p$, in 
retrieving the classical regime of the quantum system. An 
outcome of this feature is that the Liapunov exponents  that 
characterize the dynamical behavior of classical and quantum 
expectation values of the same observable can indeed be very different 
from each other \cite{balle}. The following observation 
\cite{habib} is pertinent 
in this regard. In isolated quantum systems with a discrete 
energy spectrum, using unitarity and the Schwarz 
inequality, it can be established that the Liapunov exponents would 
vanish, if computed from time-series data collected over a 
sufficiently long time (which could  be much longer 
than the characteristic time 
scales in the problem), indicative of  
non-chaotic behavior. However, once 
measurement upon the system 
is included through appropriate interaction 
with an external system, the 
corresponding Liapunov exponents need not vanish. 
In the present context, the interaction between the modes 
is effectively tantamount to continual 
measurement upon either subsystem. Hence the 
dynamics  of a  subsystem, as deduced 
from time-series data 
of the corresponding variables, may show chaotic behavior, 
even if the system as a whole does not. The  
exponential instability associated with a positive 
Liapunov exponent is  indicative of the manner in which 
an initial wave packet spreads, and 
the entanglement of the system increases,
under time evolution. These aspects are  worth 
bearing in mind, inasmuch as 
systems are ultimately quantum mechanical, and  
measured data are generically the time series of 
observables.  

This work was supported in part by 
Project No. SP/S2/K-14/2000 of the
Department of Science and Technology, India.

\end{document}